%% file: main.tex
\documentclass[10pt,conference]{IEEEtran}
\IEEEoverridecommandlockouts
\usepackage{multirow}
\usepackage{graphicx}
\usepackage[table,xcdraw]{xcolor}
\usepackage[bottom]{footmisc}
\usepackage{float}
\usepackage{cite}
\usepackage[hidelinks]{hyperref}
\usepackage{bm}
\usepackage{soul}
\usepackage{amsmath,amssymb,amsfonts}
\usepackage{mathtools}
\usepackage{graphicx}
\usepackage{textcomp}
\usepackage{xcolor}
\usepackage{multirow}
\usepackage{multicol}
\usepackage{makecell}
\usepackage{booktabs}
\usepackage{subcaption}
\usepackage{siunitx}
\DeclareSIUnit\px{px}
\def\BibTeX{{\rm B\kern-.05em{\sc i\kern-.025em b}\kern-.08em
T\kern-.1667em\lower.7ex\hbox{E}\kern-.125emX}}
\usepackage{mwe}
\usepackage[utf8]{inputenc}
\usepackage[english]{babel}
\usepackage[linesnumbered,ruled,vlined]{algorithm2e}

\renewcommand{\footnotesize}{\fontsize{9pt}{9pt}\selectfont}
\SetCommentSty{mycommfont}
\SetKwInput{KwInput}{Input}                
\SetKwInput{KwOutput}{Output}              
\usepackage[noend]{algpseudocode}
\usepackage[font=small,format=hang,parskip=0pt]{caption}
\usepackage[font=small,skip=0pt]{caption}
\usepackage{subcaption}
\usepackage{balance}
\usepackage{url}
\usepackage[inline]{enumitem}
\setlist[enumerate]{nosep}
\usepackage[normalem]{ulem}
\usepackage{xspace}
\usepackage{tikz}
\usetikzlibrary{calc}


\arrayrulecolor{black}
\usepackage[most]{tcolorbox}

\newtcolorbox{lesson}{
  colback=gray!8,
  colframe=gray!40,
  boxrule=0.4pt,
  arc=2pt,
  left=6pt, right=6pt, top=4pt, bottom=4pt,
  fonttitle=\bfseries,
  title=Lesson Learned
}

\begin{document}
\title{Cloud, Edge, or Split? Profiling Onboard and Split Vision-Language Model Deployment for Drone AI}
\author{
\IEEEauthorblockN{
Zoha Azimi\IEEEauthorrefmark{1},
Reza Farahani\IEEEauthorrefmark{2},
Schahram Dustdar\IEEEauthorrefmark{2},
Christian Timmerer\IEEEauthorrefmark{1}
}

\IEEEauthorblockA{
\IEEEauthorrefmark{1}Christian Doppler Laboratory ATHENA, Department of Information Technology (ITEC), University of Klagenfurt, Austria
}

\IEEEauthorblockA{
\IEEEauthorrefmark{2}Distributed Systems Group (DSG), TU Wien, Austria
}
}
\maketitle
\input{1-Abstract}

\input{2-Introduction}

\input{3-Background}

\input{4-RelatedWork}

\input{5-Setup}
\input{6-Evaluation}

\input{7-Conclusion}
\balance

\section*{Acknowledgment}
This work was supported by the Austrian FFG EdgeAI-Drone project, the EU ENFIELD project, the Austrian Federal Ministry for Digital and Economic Affairs, the National Foundation for Research, Technology and Development, and the Christian Doppler Research Association. Christian Doppler Laboratory ATHENA: \url{https://athena.itec.aau.at/}.

\bibliographystyle{./bibliography/IEEEtran}
\bibliography{./bibliography/IEEEabrv}
\vspace{12pt}
\end{document}

%% file: 1-Abstract.tex
\begin{abstract}
Vision-Language Models (VLMs) enable edge devices like unmanned aerial vehicles (UAVs) to interpret visual observations and reason about complex environments using natural-language instructions. However, their practical deployment remains challenging as onboard inference is constrained by limited computational, memory, and energy resources, whereas cloud-based inference introduces communication latency, bandwidth overhead, and dependence on network connectivity. To address these limitations, split computing offers a promising alternative by partitioning VLM inference between the resource-constrained UAVs and more capable remote servers. However, the performance trade-offs among fully onboard, cloud-based, and split-computing architectures for lightweight VLMs have not yet been systematically profiled. This paper benchmarks these three deployment paradigms using \textit{SmolVLM-256M} as a representative lightweight VLM. We quantify their inference latency, computational resource utilization, communication overhead, and energy consumption across varying image resolutions and network conditions. Our results show that no deployment strategy is universally optimal; instead, the preferred strategy depends on the interaction between network conditions and input image resolution.
\end{abstract}
\begin{IEEEkeywords}
Vision-Language Models (VLMs), Unmanned Aerial Vehicles (UAVs), Split Computing, Onboard Inference, Edge Computing, Drone AI.
\end{IEEEkeywords}

%% file: 2-Introduction.tex
\section{Introduction}
\label{sec:Introduction }
Vision-Language Models (VLMs) integrate visual perception with natural-language reasoning, enabling systems to interpret complex scenes and answer open-ended queries beyond predefined perception tasks. For Unmanned Aerial Vehicles (UAVs), commonly known as drones, this multimodal capability shifts aerial perception from rigid, task-specific pipelines toward flexible, interactive scene understanding~\cite{chen2026vision, sharshar2025vision}. Consequently, VLMs are gaining attraction in autonomous navigation~\cite{liu2023aerialvln}, disaster response~\cite{yaqoot2025uav}, and infrastructure inspection~\cite{chen2025bridge}, where semantic reasoning and intuitive human-UAV interaction are critical~\cite{bhattacharjya2025avery,cai2025flightgpt, farahani2026towards}.
\begin{figure}
    \centering
    \includegraphics[width=1\linewidth]{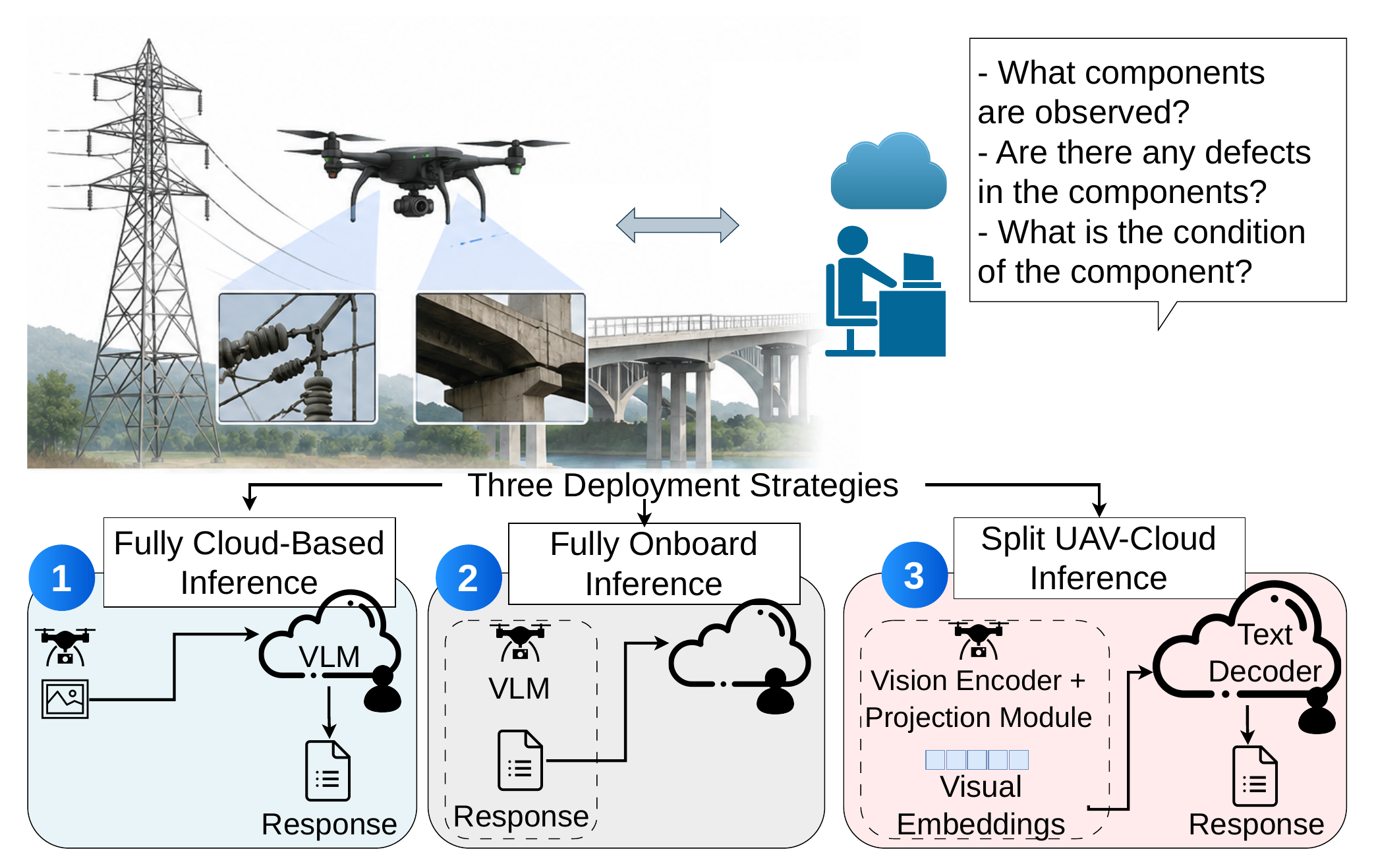}
    \caption{Overview of three VLM inference deployment strategies for UAVs: fully cloud-based, fully onboard, and split UAV-cloud inference.}
    \label{fig:deployment_strategies}
\end{figure}
Despite these capabilities, deploying VLMs on UAVs remains challenging because of their limited onboard computing, memory, energy, and communication resources.

Fig.~\ref{fig:deployment_strategies} shows three representative VLM deployment strategies: \emph{1) fully cloud-based inference}, \emph{2) fully onboard inference}, and \emph{3) split UAV–cloud inference}, each involving distinct trade-offs in onboard resource utilization, communication overhead, inference latency, and energy consumption. 

\subsubsection{Fully cloud-based inference strategy} The UAV transmits captured images to a remote cloud server, where the entire VLM inference pipeline is executed~\cite{zhang2026multimodal,chen2025bridge}. This strategy leverages abundant cloud resources to support the execution of computationally demanding VLMs~\cite{farahani2023towards}. However, image-upload delay contributes directly to end-to-end latency, making performance highly dependent on the availability, bandwidth, and stability of the UAV-cloud connection.

\subsubsection{Fully onboard inference strategy} The entire VLM pipeline is executed using the UAV's onboard edge resources~\cite{jia2025aeriaiclip,samma2026navclip,lee2025bringing}, eliminating inference-related communication overhead, preserving data locality, and enabling operation without network connectivity. However, even lightweight VLMs can impose substantial computational, memory, and energy demands on embedded hardware, potentially increasing inference latency and shortening flight endurance.

\subsubsection{Split UAV-cloud inference strategy} It splits the VLM pipeline between the UAV and a remote server~\cite{li2025distributed}, trading onboard resource usage against communication overhead and latency. As illustrated in Fig.~\ref{fig:deployment_strategies}, the UAV executes the vision encoder locally and transmits the resulting visual embeddings (i.e., latent feature tensors encoding the image content) to the cloud, where the language decoder generates the final response. Compared with fully cloud-based inference, this strategy can reduce communication volume when the embeddings are smaller than the original images. Compared with fully onboard inference, it offloads the language decoder, reducing onboard computation and memory requirements. However, its effectiveness depends jointly on the input resolution, embedding size, onboard encoder performance, and the bandwidth, latency, and stability of the UAV-cloud connection.

Despite prior work on all three strategies, the conditions under which each is preferable for lightweight VLM inference on resource-constrained UAVs remain unclear. Existing studies do not systematically compare their end-to-end latency, onboard resource utilization, communication overhead, and energy consumption under realistic experimental conditions. Moreover, network bandwidth and input image resolution alter the balance between communication and computation, making deployment performance highly context-dependent. To address these gaps, this paper empirically evaluates three lightweight VLM deployment strategies for UAV systems. Using \emph{SmolVLM-256}~\cite{marafioti2025smolvlm} as a representative VLM model, we benchmark fully cloud-based, fully onboard, and split UAV-cloud inference under realistic computing and network conditions. Experiments use an NVIDIA Jetson device as the onboard computing platform, with realistic workloads from a \emph{UAV-based power line inspection} dataset~\cite{xing2023autonomous} and the \emph{CODEBRIM bridge-defect} dataset~\cite{mundt2019meta}. 
Our results reveal distinct trade-offs in end-to-end inference latency, onboard resource utilization, communication overhead, and energy consumption across network-bandwidth and image-resolution settings. No deployment strategy consistently outperforms the others; the best choice depends jointly on network bandwidth, onboard resource constraints, and input image resolution.

The paper has six sections. Section~\ref{sec:BR} describes the VLM inference pipeline and motivates the selected UAV-cloud split point before discussing the related work in Section~\ref{sec:BR:RelatedWork}. Section~\ref{sec:IMPSetup} describes the evaluation methodology and experimental setup. Section~\ref{sec:results} reports and analyzes the empirical results and derives deployment-selection guidelines. Finally, Section~\ref{sec:conclusion} concludes the paper and outlines future directions.

%% file: 3-Background.tex
\section{VLM Pipeline and Split Inference}
\label{sec:BR}
VLMs couple visual perception with natural-language generation by connecting a vision encoder to a language model. Modern lightweight VLMs typically comprise three modules: \emph{1)} a vision encoder, \emph{2)} a projection module, and \emph{3)} an autoregressive language decoder. To clarify this data flow, we annotate each stage with the corresponding tensor shapes produced by \emph{SmolVLM-256M}~\cite{marafioti2025smolvlm,smolvlmHugface}, which is used throughout our experiments.
\subsubsection{Vision encoder} \emph{SmolVLM-256M} employs the Sigmoid Loss for Language-Image Pre-Training (SigLIP) vision transformer~\cite{zhai2023sigmoid} as its vision encoder, converting each image into high-level visual features by dividing it into fixed-size patches and modeling their relationships through self-attention. To accommodate its fixed $512\times512$ input resolution, the \emph{SmolVLM-256M} processor first resizes each source image while preserving its aspect ratio such that its longest edge is \qty{2048}{\px}. The resized dimensions are then adjusted to multiples of \qty{512}{\px}, and the image is partitioned into non-overlapping $512\times512$ tiles. A $512\times512$ thumbnail of the entire image is appended to retain global context. Let $r=\mathit{short}/\mathit{long}$ denote the ratio of the shorter image side to the longer one; for example, $r=1$ for a square image and $r=0.5$ for a $2{:}1$ image. The longer edge spans four tiles, while the shorter edge spans $\lceil4r\rceil$ tiles. The total number of encoder inputs is then $T = 4\left\lceil 4r \right\rceil + 1$, where the additional input represents the global thumbnail. Under this fixed preprocessing configuration, $T$ depends on the image aspect ratio rather than its original pixel count. For example, a square image produces a $4\times4$ grid and $T=17$, whereas a $2{:}1$ image produces a $4\times2$ grid and $T=9$. Moreover, all $T$ inputs are processed as a single batch. For each $512\times512$ input, SigLIP generates $(512/16)^2=1024$ patch embeddings, each with \num{768} dimensions. Thus, for one source image, the vision encoder produces an output tensor of shape $[T,1024,768]$.
\subsubsection{Projection module} It maps the encoder features into the language model's embedding space while compressing the visual sequence. \emph{SmolVLM-256M} employs pixel shuffle, which combines each $s\times s$ block of neighboring patch embeddings into a single representation by concatenating their feature vectors along the channel dimension. This reduces the number of spatial tokens by a factor of $s^2$ while increasing the channel dimension by the same factor, without discarding feature values. With $s=4$, each tile's $32\times32\times768$ feature grid is rearranged into an $8\times8\times\num{12288}$ representation, reducing the sequence length from \num{1024} to \num{64}. A Multi-Layer Perceptron (MLP) then projects each \num{12288}-dimensional vector to the decoder's hidden size of \num{576}, producing an output tensor of shape $[T,64,576]$. Thus, each local tile or global thumbnail contributes \num{64} visual tokens, resulting in $64T$ visual tokens that replace the corresponding image placeholder tokens.
\subsubsection{Autoregressive language decoder} \emph{SmolVLM-256M} employs \emph{SmolLM2-135M}~\cite{allal2025smollm2}, a lightweight Large Language Model (LLM) with a hidden dimension of \num{576}, to process the combined visual and text embeddings and generate the response autoregressively. Let $N_{\mathrm{text}}$ denote the number of non-image tokens. Since each source image contributes $64T$ visual tokens, the decoder input length is $L_{\mathrm{in}}=N_{\mathrm{text}}+64T$. Inference comprises two phases: \emph{(i) prefill}, which processes all $L_{\mathrm{in}}$ input tokens in parallel and initializes the key-value (KV) cache, with its computational cost largely determined by the input sequence length; and \emph{(ii) decoding}, which uses the KV cache to generate one token per step, making its latency dependent on the number of generated tokens.

The tensor shapes above determine the communication payload of the selected split point. In this work, the model is partitioned after the projection module, i.e., the vision encoder and projection module execute onboard the UAV, while the language decoder executes in the cloud. A split directly after the vision encoder would transmit a feature tensor of shape $[T,1024,768]$, whereas the selected split transmits the projected visual-token tensor of shape $[T,64,576]$. 
Assuming identical numerical precision and no additional compression, this reduces the number of transmitted values by $\frac{1024\times768}{64\times576}\approx21$, reducing the communication payload by the same factor.

%% file: 4-RelatedWork.tex
\section{Related Work}
\label{sec:BR:RelatedWork}

\begin{figure}[!t]
    \centering
    \includegraphics[width=\linewidth]{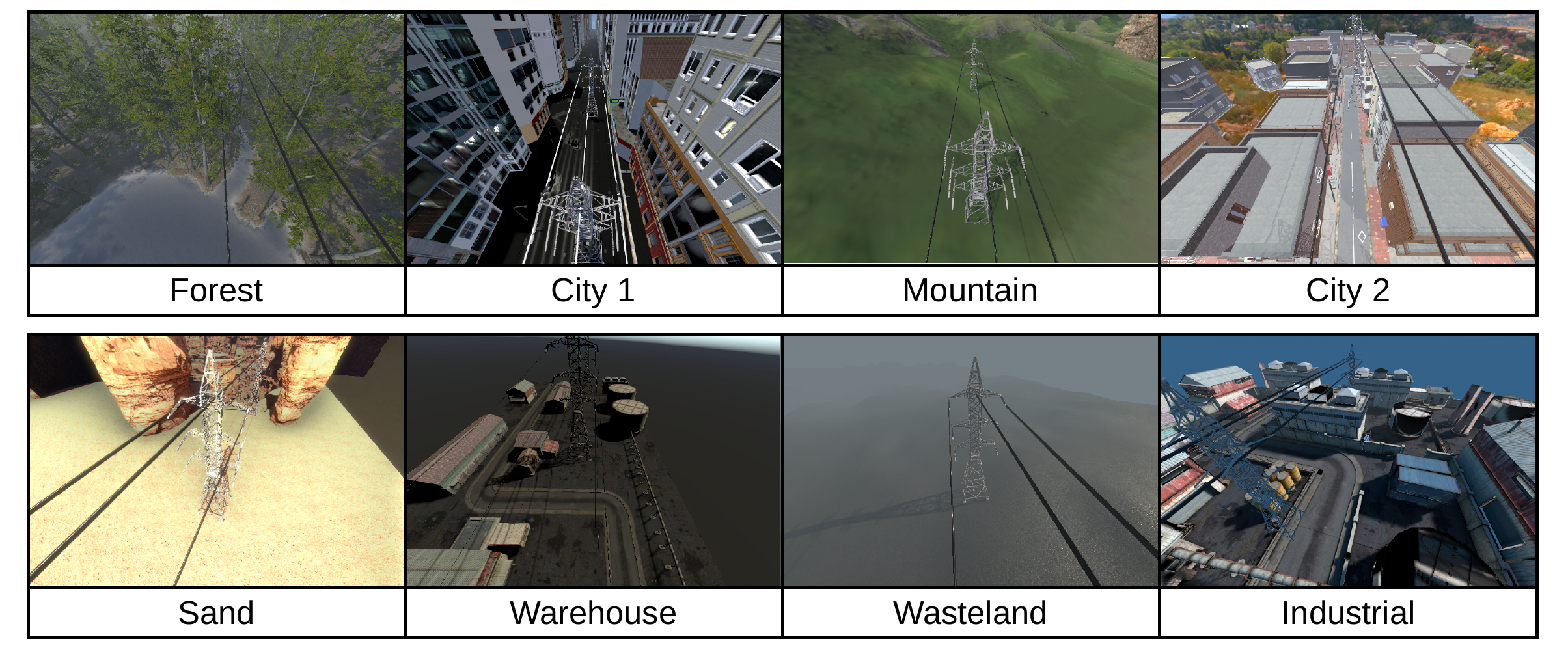}
    \caption{Representative frames from the UAV power line inspection dataset~\cite{xing2023autonomous}, covering forest, urban, mountain, sand, warehouse, wasteland, and industrial environments.}
    \label{fig:dataset}
\end{figure}
Research on UAV-based VLM systems spans both application and system-level concerns, from enabling aerial perception and reasoning to executing and distributing inference under onboard resource constraints. Thus, we categorize related work into three areas: \emph{1)} VLMs for UAV applications, \emph{2)} edge AI profiling, and \emph{3)} distributed VLM inference.
\subsubsection{VLMs for UAV applications}
Recent works have integrated VLMs into UAV-based systems to enhance perception, reasoning, and human-UAV interaction. Hu et al.~\cite{hu2025llvm} introduced LLVM-Drone, a domain-guided prompt-execution framework coupled with lightweight vision models to improve the reliability of language-driven UAV navigation. It translates natural-language instructions into executable UAV commands and visually validates their outcomes. Lelis and Dutta~\cite{lelisuavs} proposed an autonomous emergency-response framework combining UAVs, VLMs, and LLMs, where VLM-generated scene descriptions support victim detection and high-level decision-making. Chen et al.~\cite{chen2025bridge} developed a VLM-based human-UAV collaboration framework for bridge inspection that combines CLIP-based vision-language understanding with few-shot learning and prompt adaptation. Farahani et al.~\cite{farahani2026evlm} proposed EVLM, an intent-driven infrastructure monitoring framework that uses a VLM on keyframes extracted from power line video footage for defect and component detection and description.
Although these studies demonstrate the potential of VLMs for UAV autonomy, they focus primarily on application-level capabilities and do not examine the system-level trade-offs among VLM inference deployment strategies.
\subsubsection{Edge AI Profiling} The growing computational demands of foundation models have motivated profiling AI inference on resource-constrained edge platforms~\cite{azimi2026ellmpeg}. Husom et al.~\cite{husom2025sustainable} evaluated \num{28} quantized LLMs on a Raspberry Pi 4, measuring inference latency, energy consumption, and task accuracy across different quantization levels to characterize the trade-offs among model compression, execution efficiency, and output quality. Extending such analysis to multimodal models, Zhan et al.~\cite{zhan2026seeing} profiled on-device VLM inference across multiple model architectures, input resolutions, and hardware platforms, including the NVIDIA RTX 3070 and Jetson Orin NX. Their analysis quantified the relationships among inference latency, output generation, and energy consumption. Although these works characterize the device-level behavior of foundation models, they consider isolated edge devices under fixed execution configurations and do not capture the system-level trade-offs central to UAV deployment, where computation placement, wireless communication, and onboard energy constraints jointly determine performance.
\subsubsection{Distributed VLM Inference} To mitigate the computational demands of VLMs, recent works have distributed inference across edge and cloud resources. Li~et al.~\cite{li2025distributed} partitioned visual processing onto edge devices and language generation onto centralized servers, improving throughput over cloud-only inference without additional model compression. Qian~et al.~\cite{qian2025edgevlm} introduced a cloud-edge context-transfer framework that uses delayed outputs from a large cloud VLM as historical context for lightweight edge inference. Its context-replacement and visual-focus mechanisms improve responsiveness and consistency in real-time visual grounding. Zhang et al.~\cite{zhang2025vavlm} proposed an edge-cloud framework that restricts onboard processing to regions of interest to improve response accuracy. Although these works demonstrate the benefits of edge-cloud collaboration, they focus on specific distributed designs and primarily compare them with cloud-only inference. They do not quantify the collaboration cost relative to both fully onboard and fully cloud inference, leaving it unclear when split inference is preferable.
\begin{figure}
    \centering
    \includegraphics[width=1\linewidth]{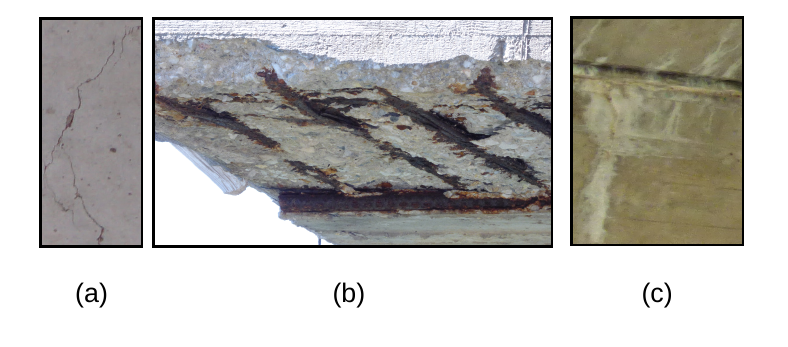}
    \caption{Representative bridge-defect images~\cite{mundt2019meta}, covering (a) crack, (b) spallation with exposed and corroded reinforcement, and (c) efflorescence.}
    \label{fig:dataset2}
\end{figure}
\begin{figure*}[t]
    \centering
    \begin{subfigure}[b]{0.3\linewidth}
        \centering
        \includegraphics[width=\linewidth]{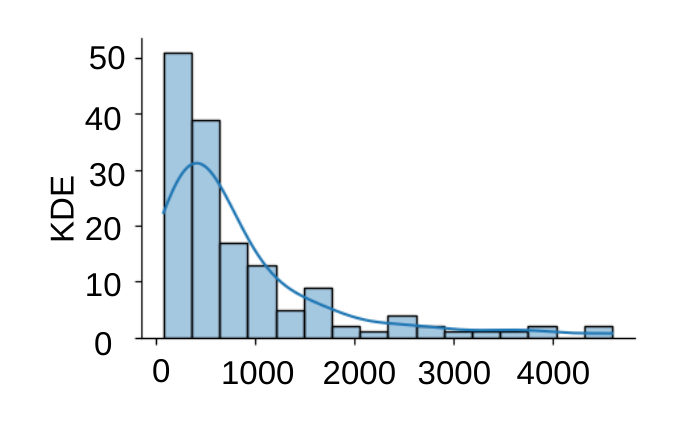}
        \caption{Image width.}
        \label{fig:dataset2_a}
    \end{subfigure}
    \hfill
    \begin{subfigure}[b]{0.3\linewidth}
        \centering
        \includegraphics[width=\linewidth]{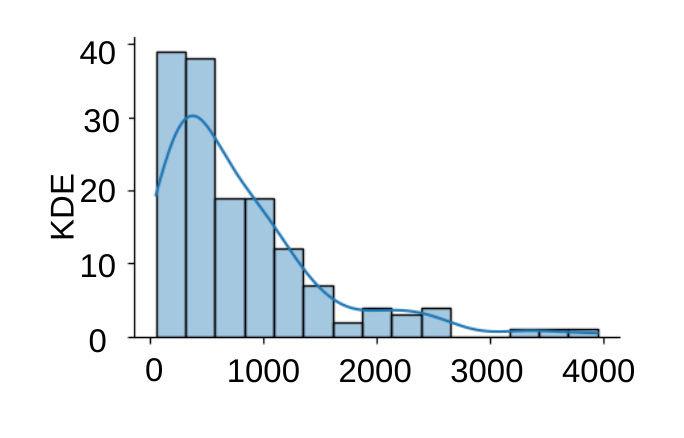}
        \caption{Image height.}
        \label{fig:dataset2_b}
    \end{subfigure}
    \hfill
    \begin{subfigure}[b]{0.3\linewidth}
        \centering
        \includegraphics[width=\linewidth]{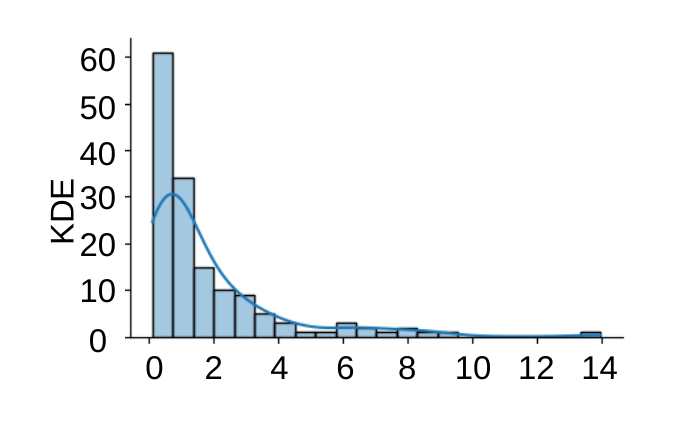}
        \caption{Image aspect ratio.}
        \label{fig:dataset2_c}
    \end{subfigure}
    \caption{Kernel density estimates (KDEs) of the spatial dimensions of the \num{150} selected CODEBRIM images.}
    \label{fig:datasetres}
\end{figure*}

%% file: 5-Setup.tex
\section{Evaluation Methodology and Experimental Setup}
\label{sec:IMPSetup}
This section presents our methodology for profiling three deployment strategies for VLM inference. It defines the deployment configurations and evaluation objectives, details the hardware specifications, datasets, inspection prompts, network conditions, and measurement metrics.
\subsection{VLM model and preprocessing configuration}
All experiments use \emph{SmolVLM-256M}~\cite{marafioti2025smolvlm}, a lightweight VLM with \num{256} million parameters. We retain its default image-processor settings, \texttt{size=\{"longest\_edge": 2048\}} and \texttt{max\_image\_size=512}, and hold them constant across all deployment configurations, ensuring that observed performance differences arise from deployment placement and experimental conditions rather than changes to model preprocessing.
\subsection{Deployment configurations}
We evaluate three VLM deployment configurations spanning fully cloud-based, fully onboard, and split UAV-cloud inference, as shown in Fig.~\ref{fig:deployment_strategies}:
\subsubsection{Fully cloud-based inference} The UAV captures the image and transmits the image-query pair to the cloud server, where the entire VLM pipeline executes.
\subsubsection{Fully onboard inference} The entire VLM pipeline executes on the UAV edge computing platform without inference-related cloud communication. 
\subsubsection{Split UAV-cloud inference} The vision encoder and projection module execute onboard the UAV, and the resulting visual tokens are transmitted to the cloud server, where the language decoder generates the response.
\subsection{Evaluation objectives}\label{sec:obj}
To compare the three deployment configurations, our evaluation is structured around the following objectives:
\subsubsection{Deployment characterization} We quantify the inference latency, CPU and GPU utilization, memory usage, and energy consumption of each deployment configuration.
\subsubsection{Network sensitivity} We determine how variations in network throughput affect communication latency and relative performance, identifying the conditions under which each deployment configuration is preferred.
\subsubsection{Input resolution sensitivity} We quantify how input image resolution affects transmitted data volume, communication overhead, and the resulting trade-offs among the three deployment configurations.
\subsection{Edge-Cloud platforms and instrumentation}
The edge device, representing the UAV's onboard computing resources, is an NVIDIA Jetson equipped with an 8-core ARM Cortex-A78AE CPU and an NVIDIA Ampere GPU with \num{1024} CUDA cores and \num{32} Tensor Cores. We use a high-performance cloud server running Ubuntu 22.04 LTS, powered by Intel Xeon Gold 5218 processors and accelerated by an NVIDIA RTX A6000 GPU.

On the Jetson, \texttt{tegrastats}~\cite{tegra} samples CPU and GPU utilization, memory usage, and module-level power at \qty{100}{\milli\second} intervals. Per-request energy is obtained by integrating the measured power over the inference interval. On the cloud server, CPU and DRAM energy are read from RAPL~\cite{rapl} counters, while GPU energy is calculated by integrating power samples collected through PyNVML~\cite{pynvml}.
\subsection{Critical-Infrastructure Inspection Datasets}
We use two public datasets for UAV-based power line and bridge inspection tasks. The power-line dataset~\cite{xing2023autonomous} provides simulated UAV video footage of power lines and transmission towers across forest, urban, mountain, sand, warehouse, wasteland, and industrial environments. We select \num{70} frames spanning these settings, with representative examples shown in Fig.~\ref{fig:dataset}.
The concrete defect bridge image (CODEBRIM) dataset~\cite{mundt2019meta} contains \num{1590} high-resolution images depicting concrete defects, including cracks, spallation, exposed and corroded reinforcement, and efflorescence. Representative samples are shown in Fig.~\ref{fig:dataset2}. Because CODEBRIM provides defect-annotated crops with heterogeneous spatial dimensions, we select \num{150} samples to evaluate sensitivity to input resolution. Fig.~\ref{fig:datasetres} summarizes the width, height, and aspect-ratio distributions of the selected images.
\subsection{VLM inspection prompt workload}
To represent common VLM workloads in UAV-based infrastructure inspection, we define \num{18} natural-language prompts grouped into four semantic categories:
\subsubsection{Component enumeration} Identification and listing of visible infrastructure elements (e.g., ``What components are present?'').
\subsubsection{Context awareness} Assessment of environmental and operational conditions, including visibility, occlusion, and weather effects (e.g., ``Are visibility conditions degraded?'').
\subsubsection{Defect detection} Identification of structural anomalies and risk factors, e.g., rust, cracks, or broken components (e.g., ``Detect rust on metal parts'', ``Is any component damaged?'').
\subsubsection{Localization} Spatial grounding of detected components or defects within the scene (e.g., ``Where is the damaged component located?'', ``Which segment contains corrosion?'').

Each image is paired with the applicable prompts, and the same image-prompt pairs are used across all deployment configurations to ensure a consistent semantic workload.
\subsection{Network trace and bandwidth conditions}
\begin{figure}
    \centering
    \includegraphics[width=0.7\linewidth]{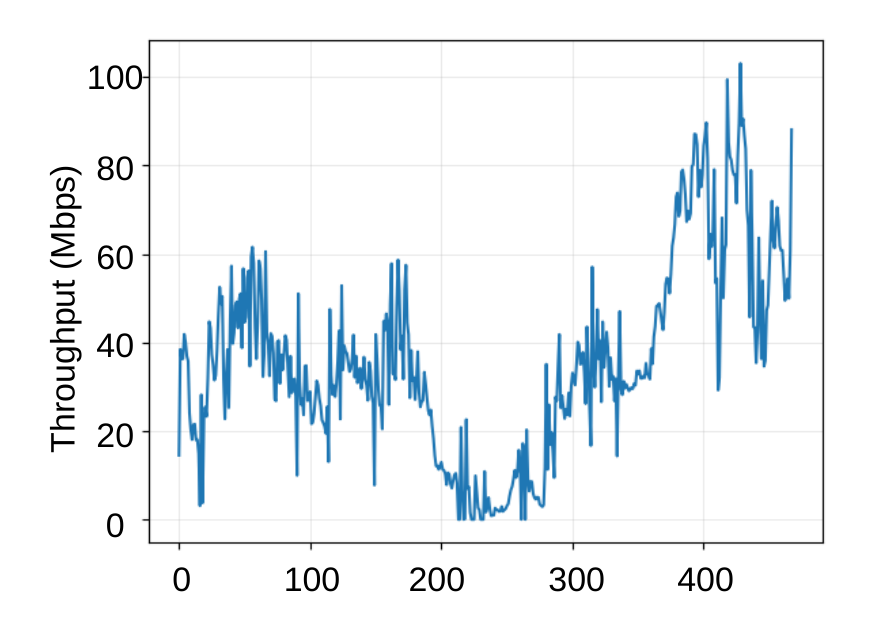}
    \caption{Time-varying throughput of the 4G/LTE network trace~\cite{van2016http}.}
    \label{fig:trace}
\end{figure}

To quantify the sensitivity of each deployment configuration to network variability, we use a \qty{500}{\second} 4G/LTE HTTP throughput trace~\cite{van2016http} to emulate time-varying UAV-cloud connectivity. As shown in Fig.~\ref{fig:trace}, the available throughput varies substantially over time. We select three representative intervals corresponding to low-, moderate-, and high-bandwidth conditions, whose mean, minimum, and maximum throughput values are summarized in Table~\ref{tab:network}.
\begin{table}[!t]
\caption{Selected bandwidth intervals and corresponding throughput statistics (Mbps).}
\label{tab:network}
\centering
\begin{tabular}{c|ccc}
\toprule
Interval & Mean & Min & Max \\ 
\midrule
Low bandwidth (\qtyrange[range-phrase=-]{220}{250}{\second}) & 2.1 & 0 & 10.86 \\
Moderate bandwidth (\qtyrange[range-phrase=-]{50}{80}{\second}) & 43.5 & 26.7 & 61.50 \\
High bandwidth (\qtyrange[range-phrase=-]{380}{480}{\second}) & 65 & 34.6 & 103.03 \\
\bottomrule
\end{tabular}
\end{table}
\subsection{Evaluation Metrics}
We compare different strategies using inference latency, communication overhead, resource utilization, and energy consumption. Unless stated otherwise, results are reported as the mean, and the standard deviation is reported across distinct image-prompt pairs.
\subsubsection{Inference latency} The execution time required to process an image-prompt pair and generate a response, excluding network transfer.
\subsubsection{Communication overhead} The transmitted payload size and corresponding transfer latency under the bandwidth conditions in Table~\ref{tab:network}.
\subsubsection{Resource utilization} The CPU and GPU utilization and peak memory usage of the participating compute platforms during inference.
\subsubsection{Energy consumption} The energy consumed by the participating compute platforms per request, obtained by integrating measured power over the inference interval.

%% file: 6-Evaluation.tex
\section{Empirical Results and Analysis}
\label{sec:results}
This section analyzes the empirical results according to the objectives defined in Section~\ref{sec:obj}. It characterizes the latency, resource, memory, and energy costs of each deployment configuration. It also evaluates sensitivity to network conditions and visual-input characteristics and derives technical guidelines for deployment selection.
\subsection{Deployment characterization}
\subsubsection{Compute latency}\label{sss:inf}
\begin{figure}
    \centering
    \includegraphics[width=0.7\linewidth]{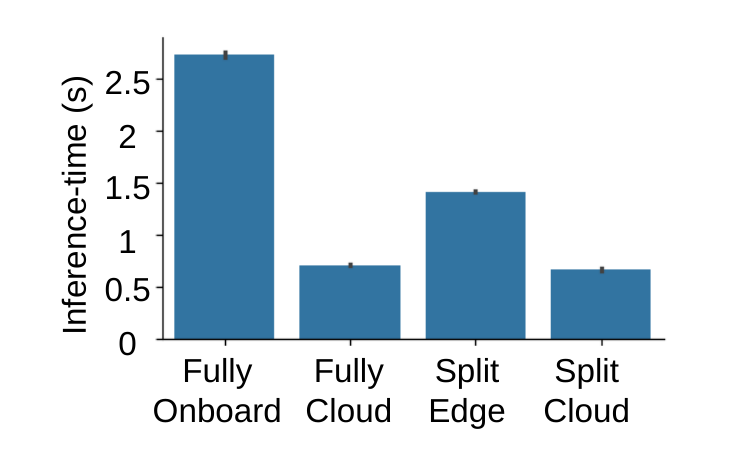}
    \caption{Average inference time per strategy.}
    \label{fig:comp_latency}
\end{figure}

Fig.~\ref{fig:comp_latency} reports the average inference time of fully cloud-based and fully onboard inference, together with the onboard and cloud stages of split UAV-cloud inference. These measurements exclude network-transfer latency and therefore isolate the execution time of each model stage.
Fully onboard inference requires approximately \qty{2.76}{\second} per request while executing the complete pipeline on the cloud reduces this latency to approximately \qty{0.72}{\second}, corresponding to a $3.8\times$ speedup due to the substantially greater computational capacity of the server. Under split inference, the onboard vision encoder and projection module require approximately \qty{1.41}{\second}, while cloud-based language decoding requires approximately \qty{0.68}{\second}. The resulting aggregate compute latency is approximately \qty{2.09}{\second}, which is \qty{24}{\percent} lower than fully onboard execution but remains substantially higher than fully cloud-based execution. These results show that offloading only the language decoder reduces the UAV's computational workload, but the onboard vision pipeline remains the dominant component of split-inference latency.
\subsubsection{Computational resource utilization and memory footprint}
\label{sec:resources}
\begin{figure}[!t]
    \centering
    \includegraphics[width=0.8\linewidth]{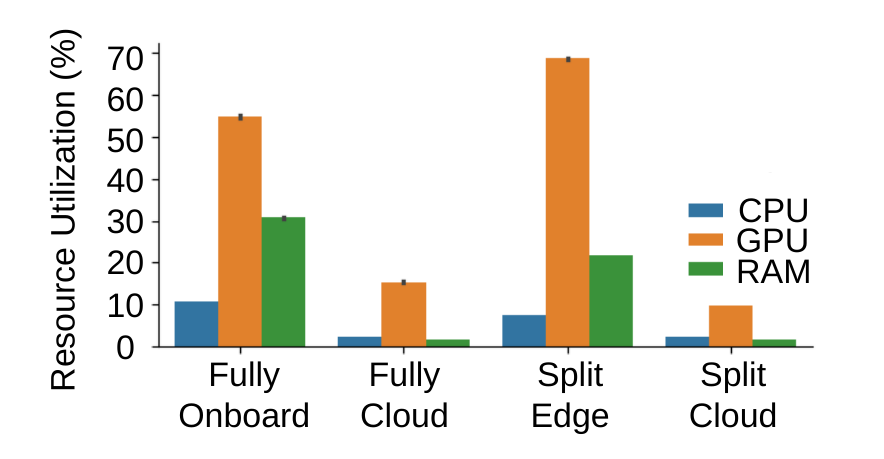}
    \caption{Average CPU, GPU, and RAM utilization across the deployment configurations and split-inference stages.}
    \label{fig:resources}
\end{figure}

Fig.~\ref{fig:resources}  compares the mean CPU, GPU, and RAM utilization of each deployment strategy. Fully onboard inference utilizes approximately \qty{11}{\percent} of the CPU, \qty{55}{\percent} of the GPU, and \qty{31}{\percent} of system memory. In contrast, fully cloud-based inference uses approximately \qty{2}{\percent} of the server CPU, \qty{15}{\percent} of its GPU, and \qty{2}{\percent} of its memory, reflecting the substantially greater capacity of the cloud platform. Moreover, the onboard stage of split inference shows approximately \qty{69}{\percent} mean GPU utilization because its measurement interval contains only the compute-intensive vision encoder and projection module. Its lower RAM utilization of approximately \qty{22}{\percent}, however, reflects the removal of the language decoder from the UAV. The higher average GPU utilization of this stage, therefore, does not imply greater total computation; rather, it results from concentrating execution on the vision pipeline.

\begin{figure}[!t]
    \centering
    \includegraphics[width=0.8\linewidth]{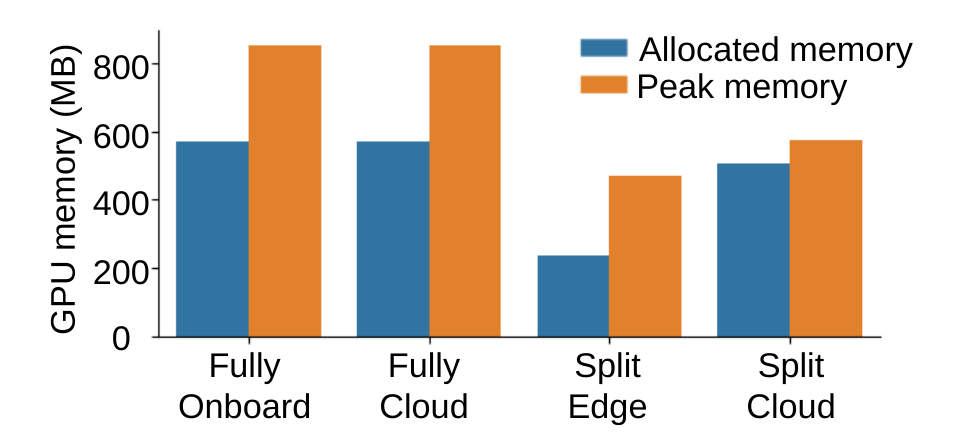}
    \caption{Average allocated and peak GPU memory across the deployment configurations and split-inference stages.}
    \label{fig:memory}
\end{figure}

The GPU-memory measurements in Fig.~\ref{fig:memory} further demonstrate the effect of model partitioning. Fully onboard execution allocates approximately \qty{572}{\mega\byte} of GPU memory and reaches a peak of approximately \qty{854}{\mega\byte}, while the onboard stage of split inference requires only approximately \qty{236}{\mega\byte} of allocated memory and \qty{472}{\mega\byte} of peak memory. Partitioning after the projection module therefore reduces onboard allocated and peak GPU memory by approximately \qty{59}{\percent} and \qty{45}{\percent}, respectively. The cloud stage of split inference also has a smaller memory footprint than executing the complete model remotely because it loads only the language decoder and its associated state. These reductions are important for UAV platforms, where GPU memory is shared with other perception, navigation, and control workloads.
\subsubsection{Energy consumption}
\begin{figure}
    \centering
    \includegraphics[width=0.7\linewidth]{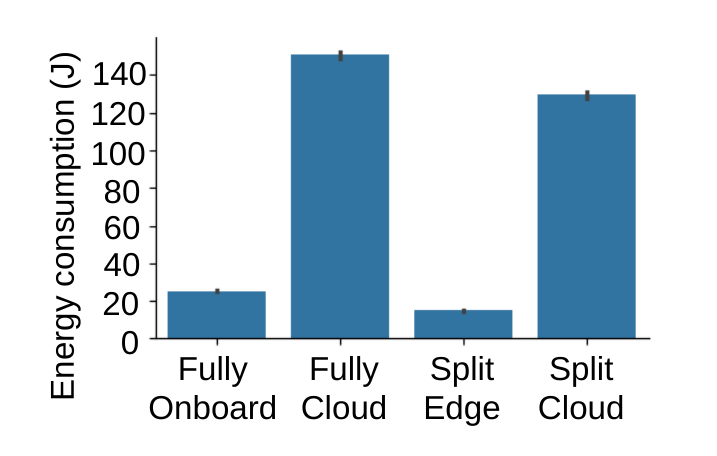}
    \caption{Average compute energy consumption across the deployment configurations and split-inference stages.}
    \label{fig:energy}
\end{figure}

Fig.~\ref{fig:energy} reports the compute energy consumed per request at each execution location. Fully onboard inference consumes approximately \qty{26}{\joule} from the UAV's onboard processing. Under split inference, the onboard vision encoder and projection module consume approximately \qty{15}{\joule}, reducing the inference-related UAV energy consumption by approximately \qty{42}{\percent}.
The cloud server consumes substantially more energy per request because it operates a higher-power CPU-GPU platform. Fully cloud-based execution consumes approximately \qty{150}{\joule}, while the cloud decoder stage of split inference consumes approximately \qty{130}{\joule}. For split inference, the total compute energy is the sum of its onboard and cloud components; however, only the onboard component draws from the UAV battery and directly affects flight endurance. Split inference consequently transfers part of the energy cost from the flight platform to fixed cloud infrastructure.
\begin{table*}[!t]
\caption{Mean transmitted payload and transfer latency under the selected bandwidth conditions.}
\label{tab:communication}
\centering
\begin{tabular}{c|c|c|c|c|c}
\toprule
\multirow{2}{*}{Dep. Strategy} & \multirow{2}{*}{Trans. data}  & \multirow{2}{*}{Trans. Size (KB)} & Trans. latency (s) & Trans. latency (s) & Trans. latency (s) \\
& & & (Low BW) & (Moderate BW) & (High BW) \\
\midrule
Fully Onboard & Text & 10  & 0.04 & 0.002 & 0.001\\
Full Cloud & JPEG-compressed image & 821.45 & 3.2 & 0.155 & 0.104\\
Split & Quantized visual embeddings & 375.84 & 1.47 & 0.071& 0.047\\
\bottomrule
\end{tabular}
\end{table*}
\begin{table}[t]
\centering
\caption{Mean end-to-end latency (s) under the selected bandwidth conditions.}
\label{tab:e2e_latency}
\begin{tabular}{lccc}
\toprule
Dep. Strategy & Low BW & Moderate BW & High BW \\

\midrule
Fully Onboard  & 2.77 & 2.73 & 2.73 \\
Fully Cloud & 3.87 & 0.83 & 0.77 \\
Split      & 3.59 & 2.19 & 2.17 \\
\bottomrule
\end{tabular}
\end{table}
\begin{figure*}[!t]
    \centering
    \includegraphics[width=0.8\linewidth]{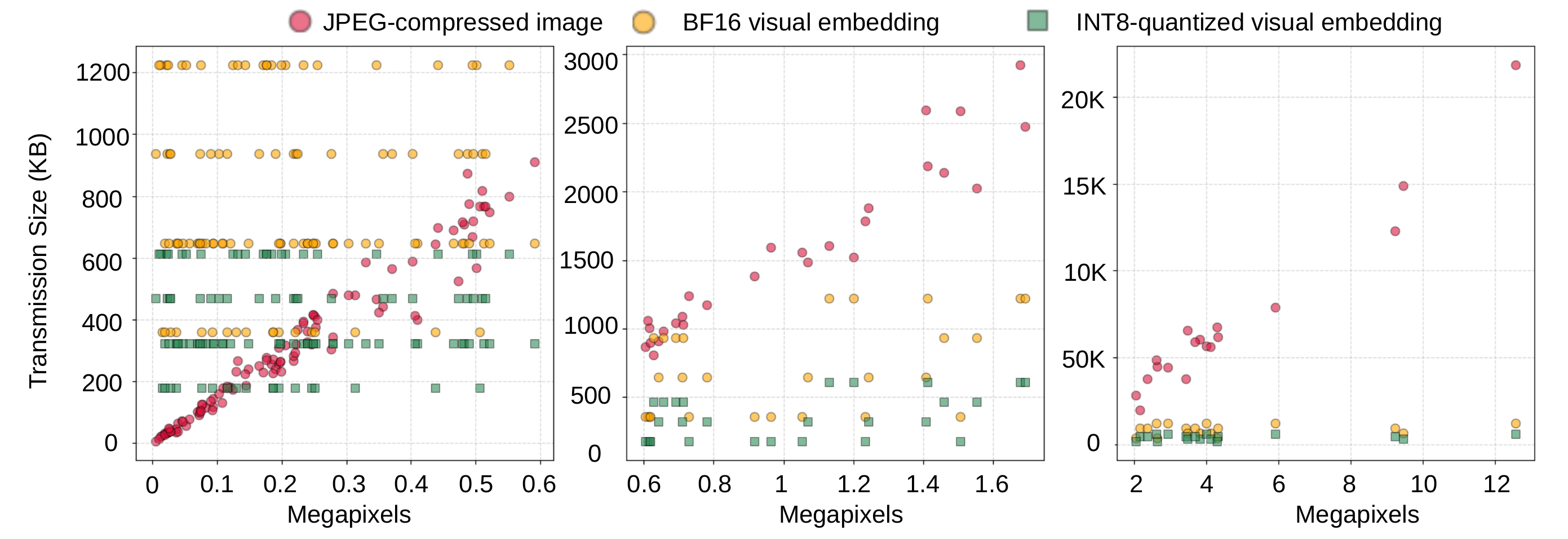}
    \caption{Transmission payload vs. source-image resolution for JPEG-compressed images and BF16 and INT8 projected visual tokens.}
    \label{fig:resImpact}
\end{figure*}
\subsection{Network sensitivity}
\subsubsection{Communication overhead}
Each deployment strategy imposes a different communication requirement. Fully onboard inference transmits only the generated text response, whereas fully cloud-based inference uploads the JPEG-compressed source image. Split inference instead uploads the INT8-quantized visual tokens produced by the onboard projection module. Table~\ref{tab:communication} reports the mean payload size and corresponding transfer latency under the bandwidth conditions defined in Table~\ref{tab:network}. Fully cloud-based inference transmits an average of \qty{821.45}{\kilo\byte}, compared with \qty{375.84}{\kilo\byte} for split inference and approximately \qty{10}{\kilo\byte} for fully onboard inference. For the evaluated workload, transmitting quantized visual tokens, therefore, reduces the uplink payload by approximately \qty{54}{\percent} relative to transmitting JPEG images. 

This reduction is most consequential under low-bandwidth conditions, where the transfer latency decreases from \qty{3.20}{\second} for fully cloud-based inference to \qty{1.47}{\second} for split inference. Under moderate and high bandwidth, the corresponding latencies are \qty{0.155}{\second} and \qty{0.104}{\second} for fully cloud-based inference and \qty{0.071}{\second} and \qty{0.047}{\second} for split inference. Fully onboard inference remains between \qty{0.001}{\second} and \qty{0.040}{\second} because it communicates only a short text response.
These results establish a clear sensitivity hierarchy, where fully onboard inference is nearly insensitive to bandwidth, fully cloud-based inference is the most sensitive because image upload dominates communication, and split inference lies between them. However, the communication advantage of split inference is not universal; it depends on whether the projected visual-token payload is smaller than the compressed source image, as examined in the input-resolution analysis.
\subsubsection{End-to-End latency}
Communication latency alone does not determine the preferred deployment strategy. End-to-end latency encompasses onboard computation, UAV-to-cloud transfer, and remote computation. As shown in Table~\ref{tab:e2e_latency}, fully onboard inference achieves the lowest latency under the low-bandwidth condition at \qty{2.77}{\second}, compared with \qty{3.59}{\second} for split inference and \qty{3.87}{\second} for fully cloud-based inference. Its limited communication requirements thus make it the most dependable option when connectivity is constrained or intermittent. Under moderate and high bandwidth, fully cloud-based inference becomes the fastest configuration, achieving \qty{0.83}{\second} and \qty{0.77}{\second}, respectively. At these bandwidths, its faster server-side execution outweighs the image-upload cost. In addition, split inference achieves \qty{2.19}{\second} and \qty{2.17}{\second}, outperforming fully onboard inference at \qty{2.73}{\second} but remaining slower than fully cloud-based inference because it combines onboard visual processing with uplink and remote-decoding latency. The latency-optimal configuration, thus, shifts from fully onboard inference under constrained bandwidth to fully cloud-based inference when sufficient bandwidth is available. Split inference does not minimize end-to-end latency under the evaluated conditions; instead, its principal benefit is balancing lower onboard memory and energy consumption against lower communication overhead than fully cloud-based inference.
\subsection{Input resolution sensitivity}
Fig.~\ref{fig:resImpact} compares the transmission payloads of JPEG-compressed source images with the BF16 and INT8 representations of the projected visual tokens. 
Unlike the JPEG size, which generally increases with source image resolution, although the exact value also depends on image content and compressibility, the visual-token payload takes a finite set of discrete values determined by the number of enoder inputs.
Specifically, with $r=\mathit{short}/\mathit{long}\in(0,1]$ the short edge spans $\lceil4r\rceil\in{1,2,3,4}$ tiles, resulting in encoder-input counts $T\in{5,9,13,17}$, as described in Section~\ref{sec:BR}. For a projected tensor of shape $[T,64,576]$, the corresponding payloads are $S_{\mathrm{BF16}} = 64T\times576\times2 = 73.7T~\mathrm{kB}$ and $S_{\mathrm{INT8}} = 64T\times576 = 36.9T~\mathrm{kB}$.
Thus, BF16 produces four possible payload levels of approximately ${370,660,960,1250}$~kB, while INT8 reduces each of these values by half.
For low-resolution inputs (below \num{0.2} megapixel), a compressed JPEG image can be smaller than either visual-token representation, limiting the communication benefit of split inference. As resolution increases, however, the JPEG payload grows substantially, exceeding \qty{20}{\mega\byte} for the largest images in the evaluated dataset, while the projected visual-token payload remains bounded by \qty{1.2}{\mega\byte} (BF16) and \qty{0.6}{\mega\byte} (INT8) at $T=17$, regardless of resolution.
Split inference, thus, becomes increasingly communication-efficient for high-resolution inputs. Nevertheless, images with similar pixel counts can produce different visual-token payloads when their aspect ratios result in different tile counts.
\subsection{Deployment selection guidelines}
The preferred VLM inference deployment strategy is determined by the end-to-end critical path, not by compute or communication latency alone. Fully cloud-based inference is latency-optimal when its server-side speedup exceeds the image-upload delay; fully onboard inference is preferable when limited or unstable bandwidth makes communication dominant. Split inference does not minimize latency under the evaluated conditions, but it substantially reduces onboard memory and energy requirements while transmitting less data than fully cloud-based inference. This trade-off is input-dependent: JPEG payload size varies with image resolution and compressibility, whereas the visual-token payload depends on aspect-ratio-driven tile count and numerical precision. Therefore, fully onboard inference suits connectivity-constrained operation, fully cloud-based inference suits latency-sensitive operation with sufficient bandwidth, and split inference suits UAVs constrained primarily by memory and battery capacity.

%% file: 7-Conclusion.tex
\section{Conclusion and Future Work}
\label{sec:conclusion}
Efficient deployment of VLM inference on resource-constrained UAVs requires jointly balancing computation, communication, and energy rather than adopting a fixed deployment strategy. This paper established this through empirical profiling of fully onboard, fully cloud-based, and split UAV-cloud inference using \emph{SmolVLM-256M}. The results show that fully onboard inference is preferable under constrained connectivity, fully cloud-based inference minimizes latency when sufficient bandwidth is available, and split inference reduces UAV-side memory and energy demands at the cost of additional communication and distributed execution. The appropriate strategy, therefore, depends jointly on network throughput, onboard resources, and visual-input characteristics.
Future work will incorporate accuracy into these trade-offs and develop an adaptive runtime orchestrator that dynamically selects the deployment strategy according to network conditions, resource availability, input characteristics, and application requirements.